\documentclass[sigconf]{acmart}

\AtBeginDocument{%
  \providecommand\BibTeX{{%
    \normalfont B\kern-0.5em{\scshape i\kern-0.25em b}\kern-0.8em\TeX}}}

\usepackage[T1]{fontenc}
\usepackage[utf8]{inputenc}

\copyrightyear{2021}
\acmYear{2021}
\setcopyright{acmcopyright}\acmConference[GLSVLSI '21]{Proceedings of the Great
Lakes Symposium on VLSI 2021}{June 22--25, 2021}{Virtual Event, USA}
\acmBooktitle{Proceedings of the Great Lakes Symposium on VLSI 2021 (GLSVLSI '21),
June 22--25, 2021, Virtual Event, USA}
\acmPrice{15.00}
\acmDOI{10.1145/3453688.3461495}
\acmISBN{978-1-4503-8393-6/21/06}

\usepackage{verbatim}
\usepackage{amsmath}
\usepackage{booktabs} % For formal tables
\usepackage{algorithm}
\usepackage{algorithmicx} 
\usepackage{algcompatible}
\usepackage{algpseudocode}
\usepackage{multirow}
\usepackage{color}
\usepackage{threeparttable}
\usepackage{pifont}
\usepackage{etaremune}
\usepackage{caption}
\usepackage{subcaption}
\usepackage[export]{adjustbox}
\usepackage{tablefootnote}

\usepackage{url}
\usepackage{enumitem}

\begin{document}
\title{\textsc{IronMan}:
GNN-assisted Des\underline{i}gn Space Explo\underline{r}ati\underline{o}n in High-Level Sy\underline{n}thesis via Reinforce\underline{m}ent Le\underline{a}r\underline{n}ing}

\author{Nan Wu}
\email{nanwu@ucsb.edu}
\affiliation{%
  \institution{UC Santa Barbara}
  \city{Santa Barbara}
  \state{CA}
  \country{USA}
  \postcode{93106}
}

\author{Yuan Xie}
\email{yuanxie@ucsb.edu}
\affiliation{%
  \institution{UC Santa Barbara}
  \city{Santa Barbara}
  \state{CA}
  \country{USA}
  \postcode{93106}
}

\author{Cong Hao}
\email{callie.hao@ece.gatech.edu}
\affiliation{%
  \institution{Georgia Institute of Technology}
  \city{Atlanta}
  \state{GA}
  \country{USA}
  \postcode{30332}
}

\begin{abstract}

Despite the great success of High-Level Synthesis (HLS) tools, we observe several unresolved challenges:
1) the high-level abstraction of programming styles in HLS conceals optimization opportunities;
2) existing HLS tools do not provide flexible trade-offs among different objectives and constraints;
3) the actual quality of the resulting RTL designs is hard to predict.
To this end, we propose an end-to-end framework, \textbf{\textsc{IronMan}}.
The primary goal is to enable a flexible and automated design space exploration (DSE), which can provide either optimized solutions under user-specified constraints, or Pareto trade-offs among different objectives (e.g., resource types, area, and latency). 
%Such DSEs either require tedious manual efforts or cannot be achieved through existing HLS tools.
\textsc{IronMan} consists of three components: 
\textbf{GPP} (a highly accurate graph-neural-network-based performance  predictor), 
\textbf{RLMD} (a reinforcement-learning-based DSE engine that explores the optimized resource allocation strategy), and \textbf{CT} (a code transformer that assists RLMD and GPP by extracting data flow graphs from original HLS C/C++).
%There are three components in \textsc{IronMan}: 
%1) GPP, a highly accurate graph-neural-network-based performance and resource predictor;
%2) RLMD, a reinforcement-learning-based multi-objective DSE engine that explores the optimal resource allocation strategy, to provide Pareto solutions among different objectives;
%3) CT, a code transformer to assist RLMD and GPP, which extracts the data flow graphs from original HLS C/C++ and automatically generates synthesizable code with HLS directives.
Experimental results show that, 1) GPP achieves high prediction accuracy, reducing prediction errors of HLS tools by $10.9\times$ in resource usage and $5.7\times$ in timing; 
2) RLMD obtains optimized or Pareto solutions outperforming genetic algorithm and simulated annealing by $12.7\%$ and $12.9\%$, respectively; 
3) \textsc{IronMan} can find optimized solutions perfectly matching various DSP constraints, with $2.54\times$ fewer DSPs and up to $6\times$ shorter latency than those of HLS tools.
\textsc{IronMan} is also up to 400$\times$ faster than meta-heuristic techniques and HLS tools.
\end{abstract}


\maketitle
\pagestyle{plain}

\vspace{-5pt}
\section{Introduction}

High-Level Synthesis (HLS) benefits ASIC and FPGA design automation by enabling automated transformation from behavioral descriptions in high-level languages (C/C++, etc.) to RTL-level designs. 
In addition to widely used commercial HLS tools for FPGA~\cite{VivadoHLS} and ASIC~\cite{StratusHLS},
recent efforts focus on improving RTL design quality \cite{cong2012optimizing, zuo2013improving}, performance and resource prediction \cite{zhao2017comba, makrani2019pyramid, zhao2019machine}, design space exploration (DSE)~\cite{schafer2019high}, etc.

Despite great achievement shown by previous efforts, there are several crucial challenges unaddressed.
\textbf{1) Higher level abstractions in HLS can obstruct optimization opportunities.}
The structured HLS coding style, such as loops and function calls, hinders advanced or fined-grained performance and resource optimization. 
Meanwhile, the irregular logic, cascaded and imperfect loops in HLS programs usually require manual or complicated code transformations to improve hardware implementation performance \cite{licht2018transformations}.
Table~\ref{tab:motivation-CT} demonstrates a simple multiplication-accumulation function using a for-loop. To explore trade-offs between the DSP usage and the number of clock cycles (latency), typical ways are to use \textit{unroll} pragmas or manual loop-tiling, as line 1-4. 
However, when the loop boundary (e.g., 8) is not divisible by the DSP constraint (e.g., 3), it results in a \textit{partial unrolling} as line 4, introducing undesired latency increment (from 4 to 8) and worsening the critical path (CP) timing (from 5ns to 7.4ns). The nested loops further complicate this problem (imagine a 5-layer nested loop with a DSP constraint of 17).
Motivated by the necessity of \textit{better performance and more flexible optimization choices}, we propose a \textbf{code transformer (CT)}. 
%An example of CT is given in Fig.~\ref{fig:DFG_with_pragma} (a).
CT easily allows to use directives, such as \textit{allocation} and \textit{resource} pragmas, to conduct finer-grained DSEs for resource and performance, as line 7-11 in Table~\ref{tab:motivation-CT}. 
%Notably, using CT+resource approaches proposed in this work (line 9-11) achieves best latency (i.e., 2) within the DSP constraint (i.e., $\leq 3$) without manual efforts.
%An alternative to constrain DSP is to use \textit{\#pragma HLS allocation instance=mul} but at the cost of increased latency (line 5-8).

\begin{figure}
    \centering
    \vspace{-20pt}
    \includegraphics[width=0.45\textwidth]{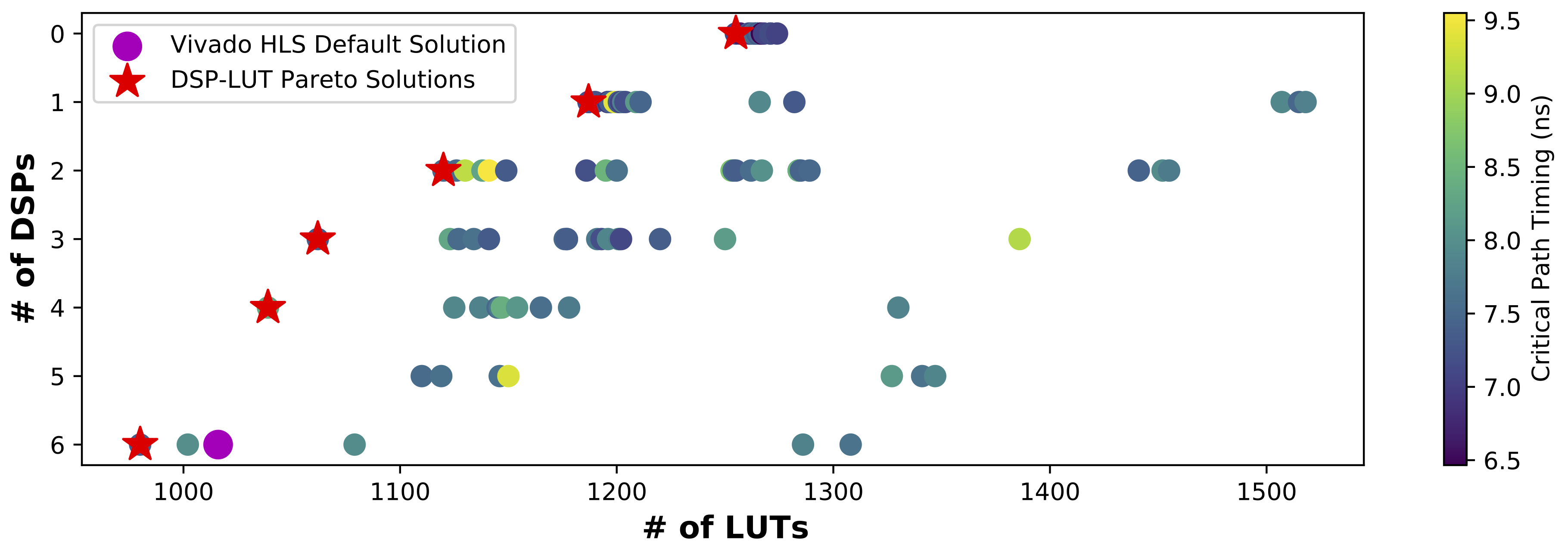}
    \vspace{-15pt}
    \caption{Pareto solutions between DSPs and LUTs on an FPGA, achieved by specifying certain multiplications using LUTs instead of DSPs. The input DFG has 200 operations.%, synthesized by Vivado HLS and implemented by Vivado.
    }
    \label{fig:motivation-DSE}
    %\vspace{-8pt}
\end{figure}

\textbf{2) HLS tools do not always provide the best solution, nor automatically provide trade-offs (Pareto solutions).} 
Existing DSE approaches as well as commercial HLS tools do not provide flexible trade-offs among different objectives and constraints (e.g., different types of resources), and they usually sacrifice design latency for less resource, or vise versa ~\cite{schafer2019high}.
In contrast, one potential alternative is to trade one type of resource for another (e.g., LUT and DSP in FPGA) while maintaining the latency, which is unexplored and only can be done through tedious manual efforts. 
An example shown in Fig.~\ref{fig:motivation-DSE} explores fine-grained trade-offs between LUTs and DSPs: 
first, the HLS default solution is not on the Pareto frontier;
second, there is a large design space for finding the Pareto solutions, and thus the \textbf{DSE for Pareto solutions is non-trivial}. 
%Given a DFG, either original or generated by CT, it requires extensive DSE to explore fine-grained trade-offs among performance, resource usage, and resource type (e.g, DSPs v.s. LUTs). 
Notably, the solution space grows exponentially even for a binary selection of DSP/LUT for each multiplication, which is further complicated by different data precisions (bitwidth).
Motivated by the necessity and difficulty of \textit{flexible and fine-grained DSE}, we propose \textbf{a deep reinforcement learning (RL) based DSE tool}, \textbf{RLMD}.

\textbf{3) The real quality of the resulting RTL designs is hard to predict, especially for irregular data paths.} 
Most existing model-based predictors target well-structured data flows, such as perfect and nested loops with high-level directives~\cite{zhao2017comba, zhong2016lin}.
As such, these predictors and high level DSE tools are \textit{not suitable for irregular logic and data paths} (especially for timing estimation).
While machine-learning-based predictions are feasible \cite{makrani2019pyramid, dai2018fast, zhao2019machine}, they often requires abundant features after design synthesis and/or implementation.
Fortunately, the inherent graph structure of data flow graphs (DFGs) provides a promising opportunity to exploit the representative power of graph neural networks (GNNs)~\cite{hamilton2017inductive, kipf2016semi, scarselli2008graph}. 
Motivated by the necessity of \textit{DSE and high-accuracy prediction for irregular data paths} and the intrinsic \textit{graph structure of DFGs}, we propose \textbf{a GNN-based HLS performance predictor}, \textbf{GPP, enabling RLMD for DSE on arbitrary DFGs}.

\begin{figure}
    \centering
    \vspace{-25pt}
    \includegraphics[width=0.45\textwidth]{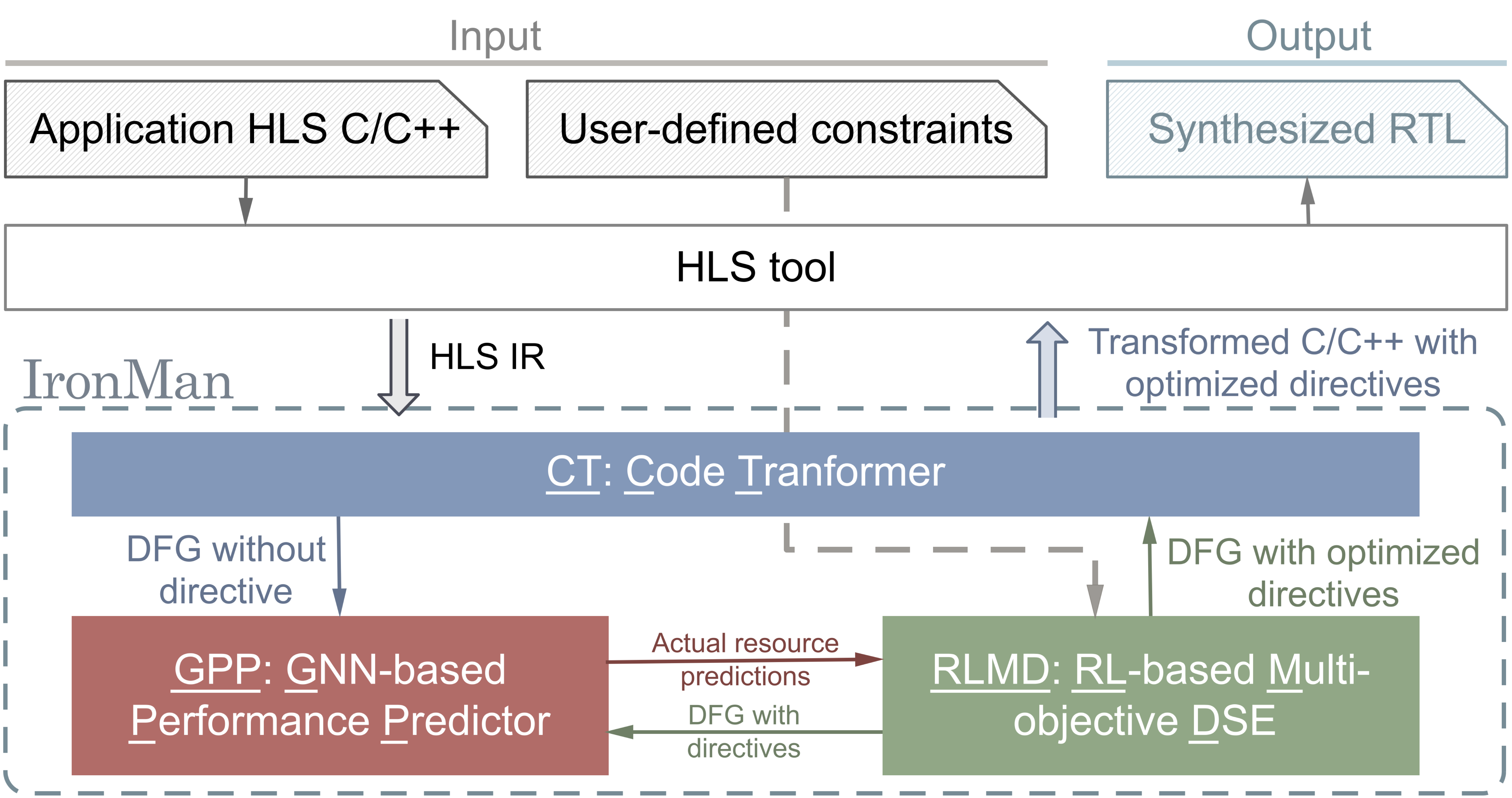}
    \vspace{-10pt}
    \caption{The overall framework, \underline{\textsc{IronMan}}.}
    \vspace{-20pt}
    \label{fig:overall}
\end{figure}

By seamlessly integrating the three aforementioned components, we propose an end-to-end framework, namely \textbf{\textsc{IronMan}}, as depicted in Fig.~\ref{fig:overall}.
\textbf{\textsc{IronMan}} has two major goals:
1) to enable a flexible and automated DSE, aiming to \textit{explore various trade-offs among different objectives such as resource types and latency};
2) to provide an accurate RTL design performance predictor, which \textit{does not require any additional features} except the original DFG, \textit{supporting both regular and irregular data paths}.
We briefly introduce the components and summarize our contributions as follows.
\vspace{-5pt}
\begin{itemize}[leftmargin=*]
    \item {
    \textbf{\underline{GPP}}: a highly accurate \textbf{\underline{G}}NN-based \textbf{\underline{p}}erformance \textbf{\underline{p}}redictor for HLS designs, including resource utilization (DSP/LUT) and critical path (CP) timing. It predicts the \textit{actual performance} after physical synthesis (placement and routing) rather than the \textit{synthesized} results by HLS tools, and can generalize to unseen DFGs. }
    \item {
    \textbf{\underline{RLMD}}: a deep \textbf{\underline{RL}}-based
    \textbf{\underline{m}}ulti-objective
    \textbf{\underline{D}}SE engine for resource allocation in HLS. Assisted by \textbf{GPP}, \textbf{RLMD} explores optimized resource allocation strategy under user-specified constraints. The objectives include minimizing resource utilization, optimizing CP timing and/or minimizing DFG computation latency. \textbf{RLMD} also provides Pareto solutions among different objectives, which are unavailable in HLS tools.
    }
    \item {
    \textbf{\underline{CT}}: a \textbf{\underline{c}}ode \textbf{\underline{t}}ransformer that extracts DFGs from original HLS C/C++ and re-generates synthesizable code with HLS directives optimized by \textbf{RLMD}. \textbf{CT} \textit{reveals concealed optimization opportunities} for achieving higher parallelism, and enables \textit{flexible and finer-grained DSE} under user-specified constraints.
    }
    \item {
    \textsc{\underline{\textbf{IronMan}}}: while each proposed component alone can contribute to the HLS community (performance prediction, DSE, code transformation), we integrate them into a framework, \textsc{IronMan}, and demonstrate the end-to-end benefits on real-world benchmarks. 
    }
    \item {
    Experimental results show that, 1) \textbf{GPP} achieves high prediction accuracy in \textit{actual} resource, reducing prediction errors of HLS tools by $10.9\times$ in resource usage and $5.7\times$ in CP timing; 2) \textbf{RLMD} obtains optimized or Pareto solutions, outperforming the genetic algorithm and simulated annealing by $12.7\%$ and $12.9\%$, respectively; 3) on real-case benchmarks, \textsc{\textbf{IronMan}} finds solutions satisfying various DSP constraints, with $2.54\times$ fewer DSPs and up to $6\times$ shorter latency than those of HLS tools, while being up to 400$\times$ faster than heuristic algorithms and HLS tools.}
\end{itemize}

\begin{table}[t]
\centering
\vspace{-15pt}
    \footnotesize
    \renewcommand{\arraystretch}{0.8}
    \setlength{\tabcolsep}{4pt}
    
\begin{threeparttable}
\small
\begin{tabular}{c|c|c|c|c|c}
\hline
\hline
\multicolumn{6}{l}{\textbf{Orig. Code: for (int i=0; i$<$8; i++) sum += a[i]$*$b[i];}} \\\hline \hline
       & \textbf{Method} & \textbf{Cycles} 
        & \textbf{DSP} & \textbf{LUTs} 
        & \textbf{CP (ns)} \\\hline
1 & Original & 17 & 1 & 75 & 4.07 \\\hline
2 & unroll (factor=8, complete) & 2 & 8 & 100 & 5.04 \\\hline
3 & unroll (factor=4) & 4 & 4 & 87 & 4.83 \\\hline
4 & unroll (factor=3) & 8 & 3 & 109 & 7.44 \\\hline
5 & unroll + allocation (limit=3) \tnote{$*$} & 4 & 6 & 168 & 8.76 \\\hline
6 & Code Transform (CT) & 2 & 8 & 100 & 5.03 \\\hline
7 & CT + allocation (limit=2) & 5 & 3 & 196 & 9.91 \\\hline
8 & CT + allocation (limit=3) & 4 & 6 & 168 & 8.54 \\\hline
9 & $\triangleright$ CT + resource (5 Mul\_LUT) & 2 & 2 & 1742 & 4.24 \\\hline
10 & $\triangleright$ CT + resource (4 Mul\_LUT) & 2 & 2 & 1741 & 4.01  \\\hline
11 & $\triangleright$ CT + resource (3 Mul\_LUT) & 2 & 3 & 1461 & 3.98 \\\hline
\hline
\end{tabular}
\begin{tablenotes}
\item[$*$] HLS pragmas do not always behave as expected.
\end{tablenotes}
%\vspace{-10pt}
\end{threeparttable}
\caption{Approaches to meeting DSP constraints (e.g., $\leq 3$).
%, with different latency, LUT usage, and critical path (CP) timing. 
This work explores \textit{CT+resource} approaches (line 9-11), which achieve the best latency under the constraint.
An alternative to constrain DSP is to use \textit{\#pragma HLS allocation instance=mul}, while increasing latency (line 7,8).
}
\label{tab:motivation-CT}
\vspace{-30pt}
\end{table}

\vspace{-8pt}
\section{Overall Framework} \label{sec:overall_framework}

\begin{figure}
\vspace{-15pt}
    \centering
    \includegraphics[width=0.44\textwidth]{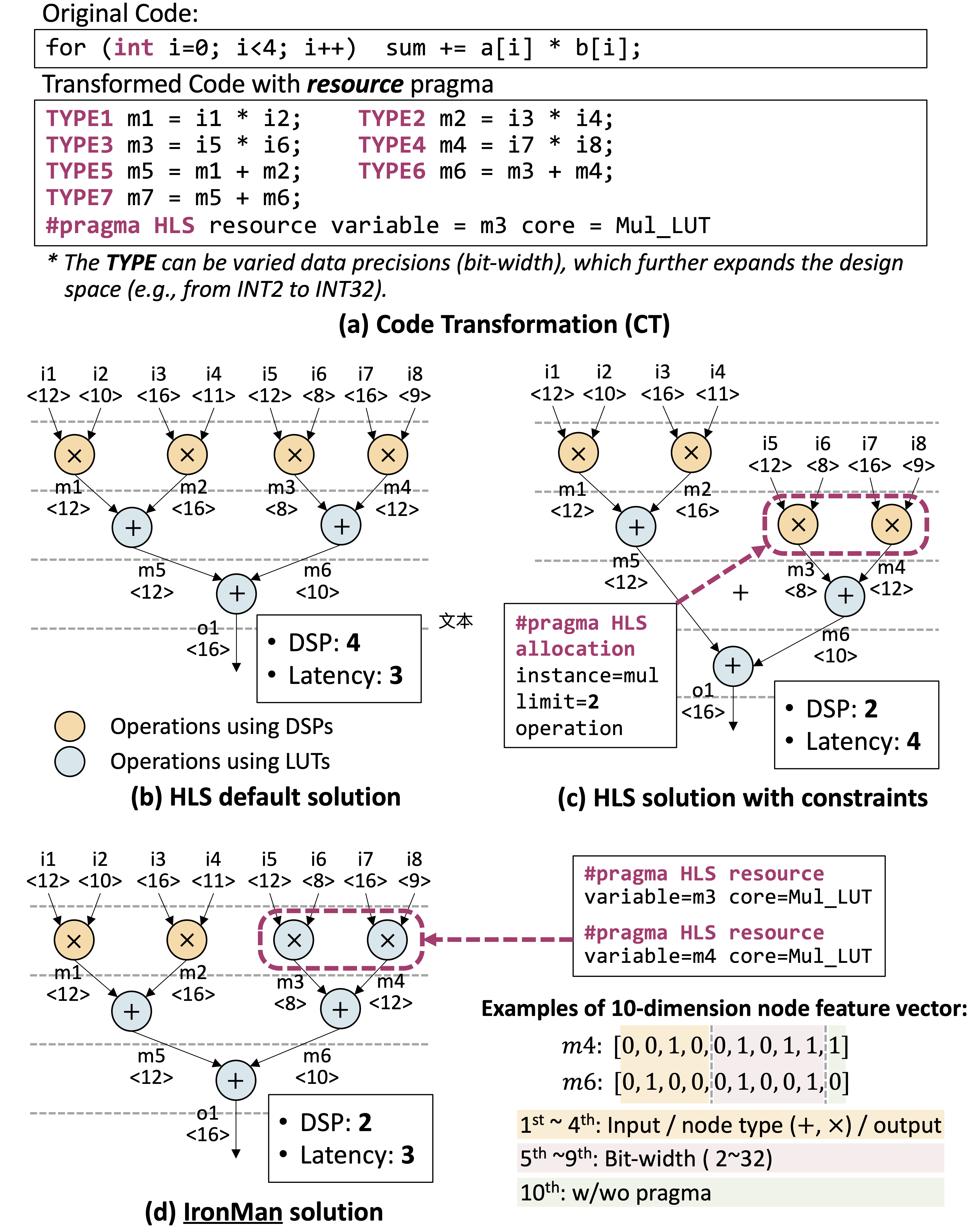}
    \vspace{-12pt}
    \caption{An example of \textsc{IronMan} solution. (a) The original HLS code and transformed code with resource pragma, indicating the importance of CT for \textsc{IronMan}. (b) HLS default solution (4 DSPs and a latency of 3); (c) HLS solution with naive constraints (2 DSPs while increasing latency from 3 to 4); (d) \textsc{IronMan} solution (2 DSPs and an unchanged latency).}
    \label{fig:DFG_with_pragma}
    \vspace{-15pt}
\end{figure}

The overall framework of \textsc{IronMan} is shown in Fig.~\ref{fig:overall}.
The inputs are HLS C/C++ code and user-specified constraints. 
The output is the re-generated code with optimized HLS directives, either meeting the user-specified constraints (e.g., resource or latency), or providing Pareto solutions among different optimization objectives.

%\textcircled{\small{1}} 
\textbf{CT} extracts the intrinsic DFGs from the intermediate representations (IRs) of HLS tools to release more optimization opportunities, and then re-generates synthesizable C/C++ code with optimized directives.
Fig.~\ref{fig:DFG_with_pragma} (a) exemplifies how CT re-generates C++ code, with the extracted DFG in (b).
Each intermediate operator may have various bit-width, e.g., $\left \langle 12 \right \rangle$ means a 12-bit data precision.

%\textcircled{\small{2}}
\textbf{GPP, a GNN-based performance predictor}, estimates the \textit{actual} resource usage after physical synthesis of DFGs.
GNNs~\cite{hamilton2017inductive, kipf2016semi, scarselli2008graph} are adopted for three reasons.
1) DFGs are graphs, which are naturally suitable for GNNs to learn the underlying information from graph structures.
2) DFGs vary in topologies and sizes, and in order to generalize predictions to unseen graphs, it is necessary to use \textit{inductive} GNNs~\cite{hamilton2017inductive} to learn fixed-size \textit{graph embeddings}.
3) \textsc{IronMan} runs inferences of trained GNN models during execution, which is orders of magnitude faster than running HLS tools.

%\textcircled{\small{3}}
\textbf{RLMD, an RL-based DSE engine}, takes DFGs, their corresponding graph embeddings, and user-specified constraints as inputs, to make endeavors for optimal resource allocation strategy.
RL is adopted for two main reasons.
1) The design space grows exponentially with the size of DFGs, different graph topologies, and various data precision. RL has been widely applied for proactive DSE in computer systems \cite{wu2021survey}, and a well pre-trained agent can generalize to new problems by minimal fine-tuning efforts.
2) By carefully defining reward functions, RL agents can achieve multi-objective optimization automatically, getting rid of manual efforts to craft useful heuristics.
An informative and well-crafted state representation will significantly benefit the learning process in RL problems, motivating the integration of GPP and RLMD.%, since the graph embeddings are naturally suitable for state representation in this problem.
Consequently, the graph embeddings enable RLMD to generalize across different DFG topologies, and GPP largely accelerates the training process of RLMD by quickly evaluating solutions generated by RLMD.

As a case study of \textsc{IronMan}, the specific problem solved is \textbf{to find a resource allocation solution that strictly meets the DSP constraint, or to find Pareto solutions between DSPs and LUTs on FPGAs, without sacrificing computation latency.}
For simplicity, the DFGs only have additions and multiplications, where RLMD decides whether to assign the directive \textit{\#pramga HLS resource core=Mul\_LUT} to each multiplication, to minimize LUTs within DSP constraints.
As shown in Fig.~\ref{fig:DFG_with_pragma}, 
%(b) is the default Vivado HLS solution; 
%(b) is the default Vivado HLS solution with 4 DSPs and a latency of 3; 
(c) naively uses \textit{\#pragma HLS allocation instance=mul limit=2} to enforce the usage of two DSPs; 
%(c) is a naive solution using \textit{\#pragma HLS allocation instance=mul limit=2} to enforce two DSPs, resulting in an increased latency from 3 to 4; 
(d) is the solution of \textsc{IronMan} with 2 DSPs and an unchanged latency of 3, using \textit{\#pragma HLS resource variable=$\langle$var$\rangle$ core=Mul\_LUT}.
Notably, such a finer-grained DSE of \textsc{IronMan} is enabled by CT.
\vspace{-5pt}
\section{Proposed GPP and RLMD} \label{sec:GPP_RLMD}
%\vspace{-4pt}

%This section first introduces the structure of GPP, and then discuss the detailed formulation of RLMD.

\begin{figure*}[ht]
    \centering
    \vspace{-15pt}
    \includegraphics[width=0.95\textwidth]{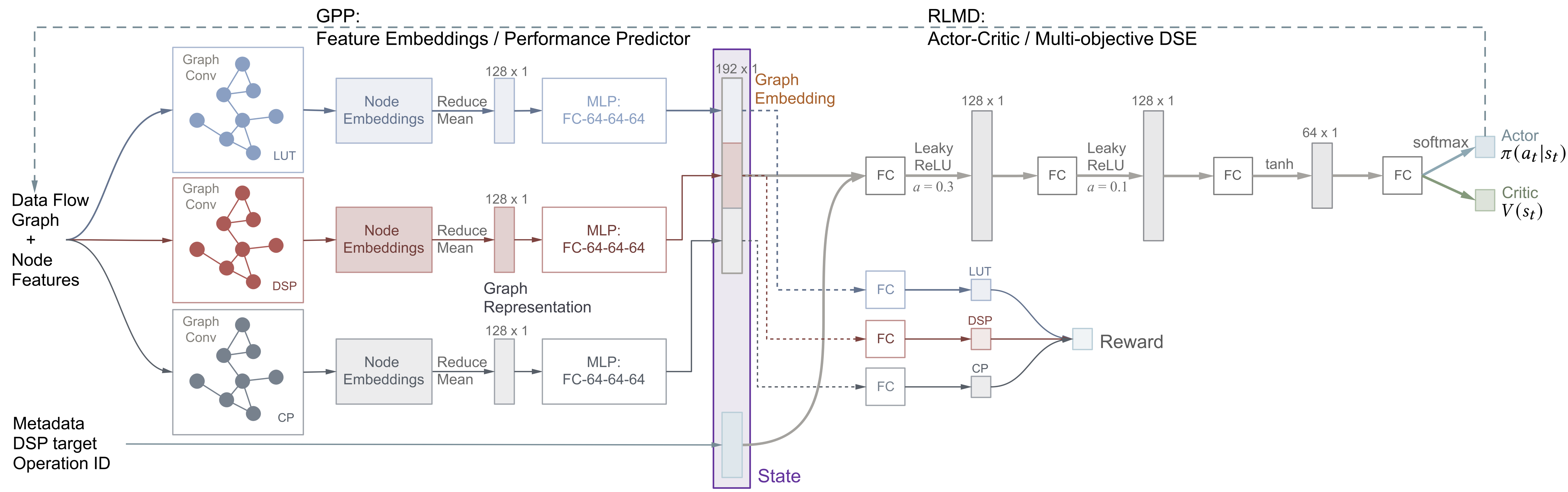}
    \vspace{-15pt}
    \caption{GPP and RLMD structure. GPP encodes information of the DFG adjacency and node features, to make predictions of LUT/DSP/CP. RLMD outputs a binary probability distribution $\pi(a_t|s_t)$ of whether to use LUTs for multiplication computation on the current node, and a scalar as the state-value function.
    %Concatenating the graph embedding provided by GPP with the meta data of the input DFG, RLMD then outputs a binary probability distribution $\pi(a_t|s_t)$ of whether to use LUTs for multiplication computation on the current node, and a scalar as the state-value function.
    }
    \vspace{-12pt}
    \label{fig:gnn_rl}
\end{figure*}

%%%%%%%%%%%%%%%%%%%%%%%%%%%%%%%%%%%%%%%%%%%%%%%%
%%%%%%%%%%%%%%%%%%%%%%%%%%%%%%%%%%%%%%%%%%%%%%%%
%\vspace{-10pt}
\subsection{GPP}
%\vspace{-4pt}

%To accurately predict physical synthesis performance and generalize to previously unseen DFGs, we apply GNNs in GPP to embed information from DFGs.
%GNNs~\cite{scarselli2008graph} adapt neural networks to process graphical data.
The key role of a GNN is to extract adequate information of node types, graph topology and connectivity within a large DFG, and encode the information into low-dimension vector representations that can be used for downstream tasks.

\textbf{Node Feature Vector.}
In a DFG, each node is encoded into a 10-dimension node feature vector,
as the example shown in Fig.~\ref{fig:DFG_with_pragma}(d).
The $1^{st}$ to $4^{th}$ dimension use one-hot representations to encode the node types, including input nodes, intermediate nodes/operations (additions and multiplications), and output nodes.
The $5^{st}$ to $9^{th}$ dimension encode the data precision of an intermediate operation, which in this work ranges from INT2 to INT32. We use a binary representation to encode the precision minus one, so the bit-width can be expressed in 5 bits.
The $10^{th}$ dimension indicates whether an HLS directive \textit{\#pramga HLS resource} is applied to this node. 
Note that such an encoding scheme can be easily extended to support more types of nodes/operations or pragmas.

\textbf{Graph Embedding.}
We employ three GNN models of the same structure to separately predict LUT/DSP usage and CP timing, as illustrated in the left part of Fig.~\ref{fig:gnn_rl}.
For each GNN model, the inputs are adjacency matrices and node feature matrices of DFGs.
The first two layers are graph convolutional~\cite{kipf2016semi}, with 64 and 128 units respectively and ReLu activations.
In each graph convolutional layer, the node embedding is updated by aggregating feature vectors from its neighbors, and one node can receive information from farther nodes by stacking multiple layers.
Next, the learned node embeddings are summarized by a mean pooling to create a graph representation (i.e., a $128\times 1$ vector).
This representation is then passed to a feed-forward network with three fully connected layers and leaky ReLu ($\alpha=0.1$) activations to generate a graph embedding (i.e., a $64\times 1$ vector).
The last layer is the output, involving a single unit with ReLu activation to provide the prediction result.

\textbf{Integration with RLMD.}
To integrate with RLMD, we combine the three embedding vectors that focus on different characteristics of DFGs into one graph embedding, shown as the $192\times 1$ vector in Fig.~\ref{fig:gnn_rl}.
Finally, the graph embedding vector is concatenated with the meta data of the input DFG, and passed to RLMD as its inputs.
The DFG meta data include the size of the DFG (i.e., the number of input/intermediate/output nodes and the number of edges) and the number of multiplications in this DFG.
Given predictions of LUT/DSP/CP, solutions generated by RLMD can be quickly evaluated, providing feedback to further improve the policy of RLMD.

%%%%%%%%%%%%%%%%%%%%%%%%%%%%%%%%%%%%%%%%%%%%%%%%
%%%%%%%%%%%%%%%%%%%%%%%%%%%%%%%%%%%%%%%%%%%%%%%%
\vspace{-5pt}
\subsection{RLMD}
%\vspace{-3pt}

\textbf{RL Formulation.}
The resource allocation problem in HLS, as a typical RL~\cite{sutton2018reinforcement} problem, can be formulated as a Markov Decision Process (MDP), with four key components.
\vspace{-3pt}
\begin{itemize}[leftmargin=*]
    \item States: the set of possible states. In this problem, a state can be every possible partially assigned DFGs.
    \item Actions: the set of eligible actions under a state. In this problem, given the current state and the currently considered node of the DFG, the action is whether to assign the directive to this node.
    \item State transition: given a state and an action, the probability distribution of next states.
    \item Reward: the immediate reward of taking an action in a state. In this problem, the reward is 0 for all intermediate actions, with an exception for the last action where the reward is the evaluation of the fully assigned DFG subject to user-specified constraints.
\end{itemize}
\vspace{-3pt}

Specifically, the state at time step $t$ is $s_t$, a concatenation of features including a $192\times 1$ graph embedding vector that describes the current status of the DFG, the ID of current node to assign a directive, metadata of the DFG, and the DSP constraint (either user-specified or automatically generated for Pareto solution exploration).
The action $a_t$ is a valid assignment of a directive to the $t^{th}$ node, i.e., whether to use LUTs for multiplication on this node.
The reward $r_t$ is defined as a negative weighted sum of predicted LUTs, CP timing, and the difference between predicted and target DSPs:
\vspace{-4pt}
\begin{equation}
\footnotesize
r_t = \begin{cases} -\alpha LUT_{p}-\beta |DSP_{target}-DSP_{p}|-\lambda CP_{p},  & t=T \\
0, & 0<t<T \end{cases}.
\label{eq:reward}
\end{equation}
where $\alpha$, $\beta$ and $\lambda$ are hyper-parameters.

At the initial state $s_0$, all the multiplication nodes in a DFG are unassigned.
At each time step $t$, the RL agent observes the current state $s_t$, takes an action $a_t$, receives a reward $r_{t+1}$ and arrives at a new state $s_{t+1}$. 
The nodes are assigned with directives sequentially based on their node IDs.
Given $T$ multiplication nodes in total, the final state $s_{T}$ corresponds to a DFG completely assigned with proper directives.
The goal is to maximize the expected rewards received.

\begin{figure*}[htbp]
    \centering
    \vspace{-30pt}
    \includegraphics[width=0.93\textwidth]{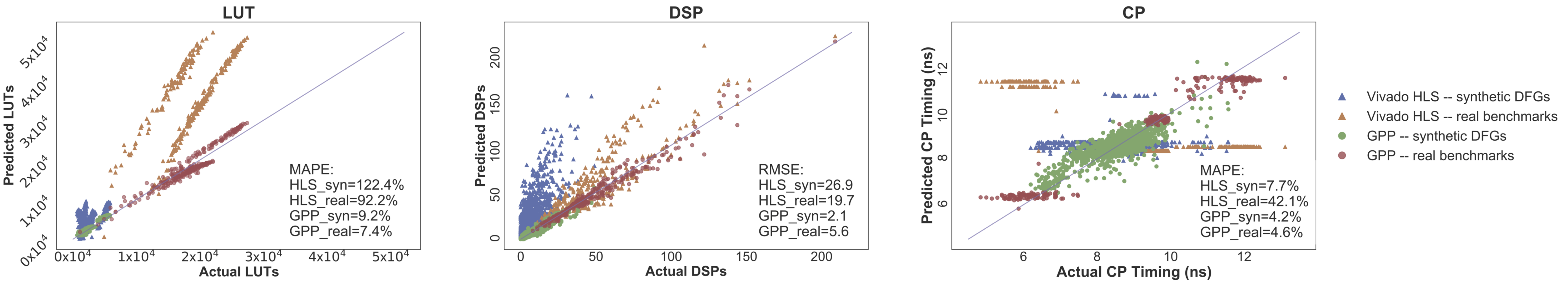}
    \vspace{-12pt}
    \caption{GPP predictions on resource utilization (LUTs and DSPs), and critical path timing (CP timing).}
    \vspace{-10pt}
    \label{fig:GPP_results}
\end{figure*}

\textbf{RLMD Training.}
We adopt the actor-critic method with Monte-Carlo learning~\cite{sutton2018reinforcement}:
the actor aims to learn an optimal policy $\pi_\theta (a_t|s_t)$ parameterized by $\theta$, which is a probability distribution of valid actions under the current state;
the critic approximates the state-value function $V(s_t)=\mathbb{E_\pi}\lbrack {\sum_{k=0}^{T-t} \gamma^kr_{t+k}}|s_t\rbrack$ by parameters $w$, which is an estimate of total rewards starting from state $s_t$ to $s_{T}$ following policy $\pi$. %, measuring the goodness of this state.
The $\gamma\in(0,1]$ is the discount factor.
As shown in the right part of Fig.~\ref{fig:gnn_rl}, there are shared parameters in the actor $\pi_\theta$ and the critic $V_w$, and for clarity we denote $\theta$ and $w$ separately.
By Monte-Carlo learning, the parameters are updated once only after one complete episode (i.e., one complete assignment process of a DFG), leading to the updates as follows:
%\vspace{-4pt}
\begin{equation}
\footnotesize
    \delta_i=\gamma ^{T-i} r_T - V_w(s_i),
    \label{eq:td_error}
\end{equation}
\vspace{-13pt}
\begin{equation}
\footnotesize
    \Delta w \propto \sum_{i=1}^{T} \delta_i \nabla_w V_w(s_i),
    \label{eq:critic}
\end{equation}
\vspace{-10pt}
\begin{equation}
\footnotesize
    \Delta \theta \propto \sum_{i=1}^{T} \delta_i \nabla_{\theta} \log \pi_{\theta}(a_i|s_i),
    \label{eq:actor}
\end{equation}
where $T$ is the total time steps in one episode.
Through repeated episodes (i.e., sequences of states, actions, and rewards), the actor learns optimized policy that will maximize cumulative rewards.

Our ultimate goal is to enable RLMD to generate higher-quality results and transfer knowledge across various DFGs as it gains experience from exploring resource allocation strategies on more and more DFGs.
Thus, we formally formulate the overall optimization objective function as:
\vspace{-6pt}
\begin{equation}
\footnotesize
    \mathcal{J}(\theta,w,G)=\frac{1}{K} \sum_{g\in G} \mathbb{E}_{g,l\sim \pi_{\theta}}
    [R_{g,l}],
\end{equation}
where $\mathcal{J}(\theta,w,G)$ measures the total expected rewards over all training DFGs.
The dataset $G$ has $K$ different DFGs, each of which is denoted as $g$.
$R_{g,l}$ is the episode reward (i.e., $r_T$ in Eq.(\ref{eq:reward})) under the resource allocation $l$ on the DFG $g$.
To get better exploration during training, we apply $\epsilon$-greedy algorithm for action selections~\cite{sutton2018reinforcement}.

\textbf{RLMD Fine-tuning.}
Given an unseen DFG, the simplest way is to directly apply the pre-trained RLMD for inference, which can generate a solution within a second.
When higher quality solutions are expected, the pre-trained RLMD can be further finetuned on this particular DFG.
The fine-tuning step provides the flexibility to trade off between a quick solution using the pre-trained RLMD (which has learned rich knowledge of resource allocation strategies on other DFGs) and a longer yet better one for a particular DFG.
\vspace{-4pt}
\section{Experiment} \label{sec:experiment_result}

\begin{figure}
    \centering
    \vspace{-3pt}
    \includegraphics[width=0.47\textwidth]{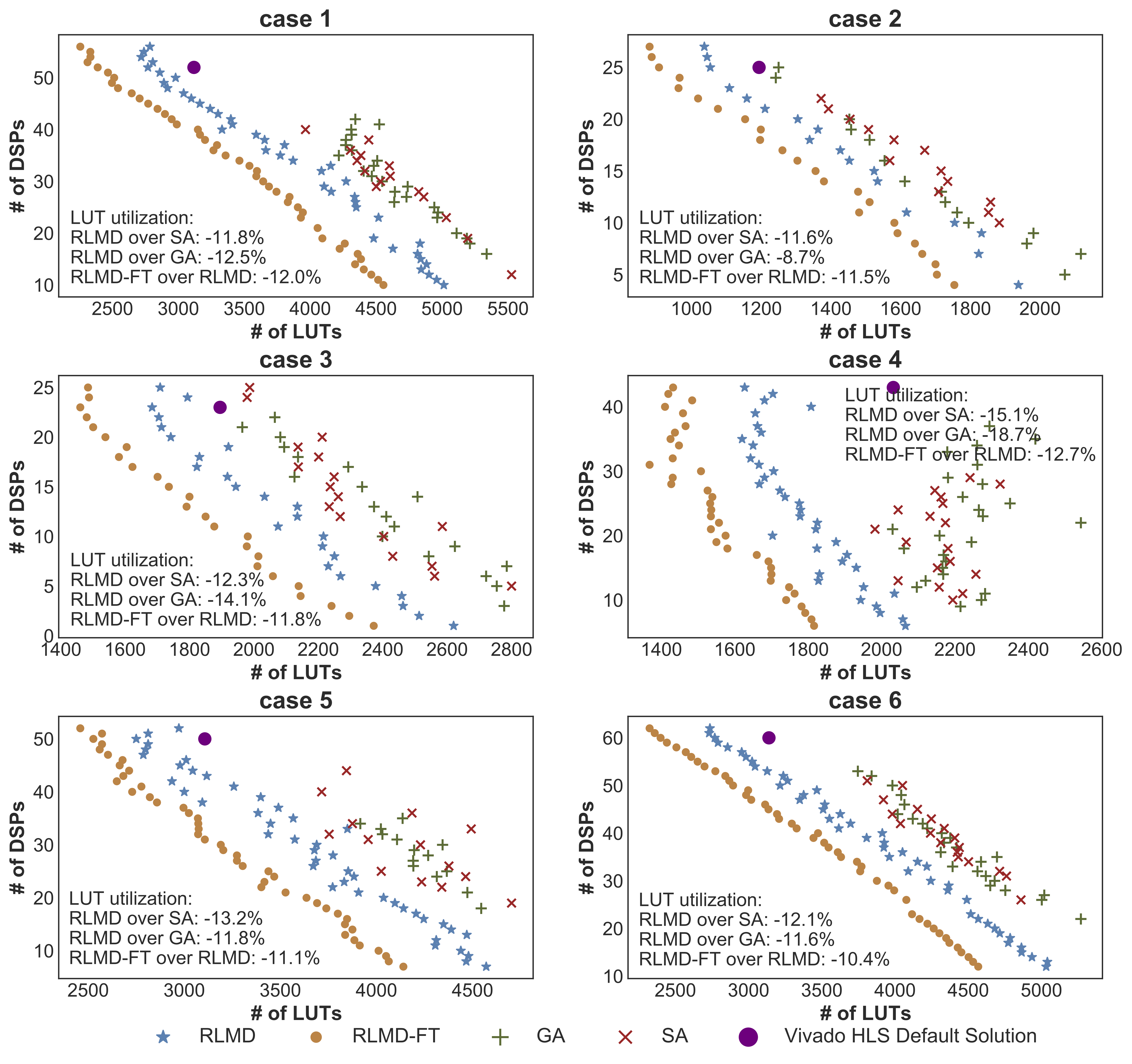}
    \vspace{-15pt}
    \caption{Pareto solutions found by RLMD, SA and GA on synthetic DFGs, with unchanged latency.}
    \vspace{-15pt}
    \label{fig:RLMD_syn_results}
\end{figure}

\begin{figure*}[h]
    \centering
    \vspace{-35pt}
    \includegraphics[width=0.96\textwidth]{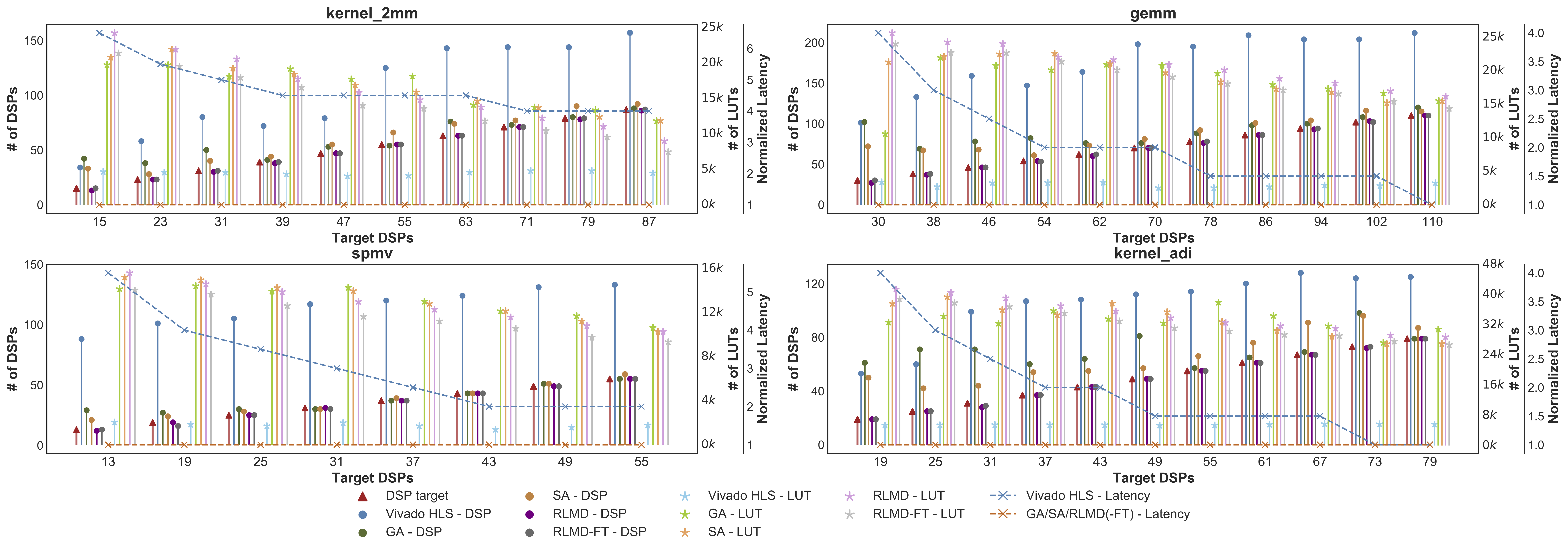}
    \vspace{-15pt}
    \caption{\textsc{IronMan} performance on real-case benchmarks. \textsc{IronMan} meets DSP constraints for 39 out of 40 cases, and meets all 40 after fine-tuning, while SA, GA and Vivado HLS meet for 2, 6, and 0 cases, respectively. %With $1.3\times$ and $1.44\times$ less DSP usage, \textsc{IronMan} after fine-tuning uses $7.7\%$ and $5.9\%$ less LUTs than SA and GA, respectively. Meanwhile, \textsc{IronMan} always maintains the shortest latency while Vivado HLS increases the latency by up to $6\times$.
    }
    \label{fig:Ironman_results}
    \vspace{-15pt}
\end{figure*}

\subsection{Experiment Setup}
\vspace{-3pt}
\textbf{Dataset Generation.}
To train GPP and RLMD, we build a dataset of both synthetic and real-case DFGs, which are generated by our CT.
For synthetic DFGs, we randomly generate 47 different topologies, each of which has 100 to 200 operations (i.e., intermediate nodes) of either multiplication or addition.
%, with random data precision between INT2 and INT16.
Upon each distinct topology, we generate 100 sets of directives, specifying a subset of multiplications to be implemented by LUTs rather than DSPs.
This makes up 4700 (i.e., 47 $\times$ 100) different synthetic DFGs.
For real-case DFGs, 8 benchmarks from MachSuite~\cite{reagen2014machsuite}, CHStone~\cite{hara2009proposal} and PolyBench/C~\cite{PolyBench} are considered:
\textit{gemm, kernel\_2mm, kernel\_durbin} (small, large), \textit{spmv, stencil3d} (small, large), and \textit{kernel\_adi}.
%The total number of training cases is 5458.
Similarly, we randomly generate 100 sets of directives per benchmark, making up 800 real-case DFGs.
The ground-truth (actual) resource usage (LUT/DSP) and CP timing are synthesized by Vivado HLS \cite{VivadoHLS} and implemented by Vivado \cite{Vivado} targeting Xilinx Ultra96 part xc7z020clg484.

\textbf{Training Process.}
To demonstrate the generalization capability of \textsc{IronMan} across different DFGs and applications, GPP and RLMD are trained on part of DFGs from the dataset and evaluated on rest of them.
The training set consists of 41 different topologies and 4 real-case benchmarks (\textit{kernel\_durbin} and \textit{stencil3d}), totally involving 4500 distinct DFGs.

GPP is trained via regression to minimize the mean squared logarithmic errors for DSPs and CP timing, and the mean absolute errors for LUTs, respectively.
The supervised learning process and GNN models enable GPP to identify features and necessary information to generalize performance prediction and graph embeddings across different DFGs. 
In terms of hyper-parameter selection, the GPP is trained over 200 epochs with batchsize as 32.
The Adam optimizer is applied with an initial learning rate of 0.01 decaying exponentially.

Once GPP is trained, it is integrated with RLMD and the training process of RLMD can be started.
To train RLMD, we provide tuples in the form of $[DFG_{index}, DSP_{target}]$ to the RL agent. 
The optimization goal is to maximize the average cumulative rewards on all of the tuples, so that the agent can learn resource allocation strategies under different DSP constraints and across different DFGs.
There have 1125 different tuples in total, and each tuple appears 8 times during the training process, amounting to 9,000 episodes.
As for fine-tuing the RLMD on a particular DFG, we additionally conduct 500 episodes of training with various target DSPs.
The parameters in RLMD are also learned by Adam optimizer with the learning rate of 0.008.
We empirically set the discount rate $\gamma=0.95$, and the exploration rate $\epsilon=0.08$ decaying exponentially.
In the reward function, we have $\alpha=0.002,\beta=5$, and $\lambda=0.02$.

% \textbf{Baselines.}
% We compare \textsc{IronMan} with three optimization methods: 1) genetic algorithm~\cite{whitley1994genetic} (denoted by GA), one prominent instance of evolutionary optimization algorithms;
% 2) simulated annealing~\cite{van1987simulated} (denoted by SA), an effective technique to approximate the global optima in an extremely large search space;
% 3) Vivado HLS, the default solutions provided by the Vivado HLS tool~\cite{VivadoHLS}.
% Solutions from different optimization methods are finally synthesized and implemented by Vivado to derive the actual performance.

\vspace{-6pt}
\subsection{Evaluation}
%\vspace{-4pt}

\textbf{Baselines.}
To evaluate GPP, we compare with the commercial tool Vivado HLS~\cite{VivadoHLS} and the learning-based performance predictor, Pyramid~\cite{makrani2019pyramid}.
To evaluate \textsc{IronMan}, we compare with genetic algorithm (GA)~\cite{whitley1994genetic},
simulated annealing (SA)~\cite{van1987simulated}, and the solutions provided by Vivado HLS~\cite{VivadoHLS}.
Notably, none of the state-of-the-art DSE methods for HLS~\cite{schafer2019high} can do such a fine-grained resource allocation as we proposed in this work, which is enabled by our CT. 
%Solutions from different optimization methods are finally synthesized and implemented by Vivado to derive the actual performance.

\textbf{GPP vs. HLS tool and Pyramid~\cite{makrani2019pyramid}.}
GPP is evaluated on both synthetic and real-case DFGs. 
Fig.~\ref{fig:GPP_results} compares GPP predictions with HLS synthesis reports regarding LUT, DSP and CP timing.
For LUT usage, the mean absolute percentage errors (MAPEs) of GPP on synthetic and real-case DFGs are $7.4\%$ and $9.2\%$, whereas the MAPEs of Vivado HLS are $122.4\%$ and $92.2\%$, respectively.
For DSP usage, the prediction accuracy is measured by root-mean-square error (RMSE) since MAPE is not applicable when the ground truth appears to be zero. 
GPP achieves 5.6 and 2.1 in RMSE for synthetic and real-case DFGs, while Vivado HLS reaches 26.9 and 19.7, respectively.
For CP timing, the MAPEs of GPP are $4.2\%$ and $4.6\%$ on synthetic and real-case DFGs, whereas the MAPEs of Vivado HLS are $7.7\%$ and $42.1\%$.
On average, GPP reduces the prediction error of Vivado HLS by $10.9\times$ in resource and $5.7\times$ in timing.

Pyramid~\cite{makrani2019pyramid} is also an ML-based framework for resource and timing prediction. 
The major difference between GPP and Pyramid is the features required for predictions.
Pyramid needs 72 features from HLS reports as inputs, which enforce the running of HLS to get VHDL designs, possibly consuming hours for large designs; 
whereas GPP can make high-accuracy predictions simply from raw DFGs (within a second).
Pyramid considers four ML models and an ensemble of these four, none of which includes graphical structure.
The reported results show that the averaged prediction error of a single ML model is $17.8\%$ for resource and $17.3\%$ for timing, with the ensemble reaching $5.5\%$ for resource and $4.1\%$ for timing.

\textbf{RLMD vs. GA/SA.}
Fig.~\ref{fig:RLMD_syn_results} compares RLMD with GA and SA regarding the Pareto solutions between LUTs and DSPs.
Obviously, RLMD outperforms GA and SA by a large margin.
Given the same number of DSPs, RLMD can find solutions reducing the LUT usage by $12.7\%$ and $12.9\%$, compared with SA and GA.
After fine-tuning, additional $11.6\%$ reduction in LUT utilization is achieved.

These promising results show great potentials of applying RL for DSE in HLS.
Through trials and interactions with GPP and user-specified constraints, RLMD is able to gradually understand which directive should be assigned to which node, and proactively learn proper resource allocation strategies by balanced exploration and exploitation.
In contrast, one underlying assumption in GA is that the offspring of two strong individuals among a population is often stronger, which is not the case in DSE for HLS problems, thus reducing its effectiveness.
Similarly, SA is a probabilistic technique and uses meta-heuristic aiming to approximate the global optima, which ignores past experiences and searches solutions to some extent hinging on randomness, thus not always reliable.

\textbf{\textsc{IronMan} vs. All.}
As depicted in Fig.~\ref{fig:Ironman_results}, \textsc{IronMan} is fully evaluated by comparing with SA, GA and Vivado HLS on four real-case benchmarks, %\textit{kernel\_2mm}, \textit{gemm}, \textit{spmv}, and \textit{kernel\_adi},
whose sizes are $2-4\times$ larger than synthetic DFGs.
To showcase \textsc{IronMan} capable to perfectly satisfy user specifications without sacrificing latency, we specify different DSP constraints within $20\%$ to $80\%$ of the maximal number of DSPs for each case.

Among four real-case benchmarks with 40 different DSP constraints in total, \textsc{IronMan} is able to meet 39 (97.5\%) of them, and can further improve to 40 (100\%) by fine-tuning.
Whereas SA, GA and Vivado HLS only meet the constraints for 2 (5\%), 6 (15\%) and 0 cases, respectively. 
Specifically, \textsc{IronMan} on average consumes 98.5\% of the targeted DSPs (improved to 99.3\% with fine-tuning), whereas those found by SA, GA and Vivado HLS use 1.29$\times$, 1.43$\times$, and 2.54$\times$ targeted DSPs, respectively. 
Not only can \textsc{IronMan} meet user-specified constraints much more accurately than its counterparts and HLS tools, but it always maintains the shortest latency whereas Vivado HLS results in an increased latency by up to $6\times$.

Reducing DSPs without sacrificing latency is at the cost of increased LUTs,
because the DSP resource is often more critical in FPGAs while LUTs are more adequate.
By perfectly satisfying DSP constraints, \textsc{IronMan} slightly increases the LUT usage by 1.2\% and 3.0\%, compared with SA and GA; 
with further fine-tuning, \textsc{IronMan} achieves additional 8.9\% LUT reduction, resulting in 7.7\% and 5.9\% lower LUT usage than SA and GA.

\textbf{Execution Time.}
During inference, i.e., being applied on real applications, \textsc{IronMan} only takes a few seconds for prediction and solution generation. 
Vivado HLS takes tens of minutes to synthesize C++ code, and up to hours to get the exact resource usage after implementation.
SA and GA take hours in average, as they struggle to closely meet the DSP constraint, and cannot generalize across different DFGs or DSP constraints. The fine-tuning for RL agent can balance between a quick solution using the pre-trained model and a longer yet better one for a particular DFG, which is optional and the number of episodes is adjustable regarding users' requirements.

\vspace{-7pt}
\section{Conclusion}
%\vspace{-3pt}
\textsc{IronMan} is an end-to-end framework, aiming to help HLS tools generate higher quality solutions under user-specified constraints, or perform more flexible DSEs to provide Pareto solutions that are not currently supported by HLS tools.
\textsc{IronMan} is equipped with a GNN-based performance predictor GPP, an RL-based DSE engine RLMD, and a code transformer. 
Independently, GPP achieves high prediction accuracy; 
RLMD obtains Pareto solutions surpassing GA and SA.
Integrated, \textsc{IronMan} is capable to find optimized solutions perfectly matching various DSP constraints, with up to 400$\times$ faster than the heuristic algorithms and HLS tools.

%Independently, GPP achieves high prediction accuracy, reducing prediction errors of HLS tools by $5.7\times$ in timing and $10.9\times$ in resource; 
%RLMD obtains Pareto solutions that outperform the GA and SA by $12.7\%$ and $12.9\%$.
%Integrated, \textsc{IronMan} is capable to find optimized solutions perfectly matching various DSP constraints, with $2.54\times$ fewer DSPs and up to $6\times$ shorter latency than those of HLS tools.
%\textsc{IronMan} is also up to 400$\times$ faster than the heuristic algorithms and HLS tools.

\bibliographystyle{ACM-Reference-Format}
\bibliography{ref}

\end{document}